\documentclass[11pt]{article} 
\usepackage{amsmath,amstext,amsbsy,amssymb,amsthm,amscd,array}
\usepackage{bm}
\usepackage{graphics,xcolor}
\usepackage{slashed}
\usepackage{cite}
\usepackage{color}
\usepackage{dsfont}
\usepackage{soul} 
\usepackage[utf8]{inputenc}
\newcommand{\Tr}{{\rm Tr} }

\newcommand{\be}{\begin{equation}}
\newcommand{\ee}{\end{equation}}
\newcommand{\w}{\wedge}

\newcommand{\ssR}{{\scriptscriptstyle{R}}}

\newcommand{\ssS}{{\scriptscriptstyle{S}}}

\newcommand{\ssD}{{\scriptscriptstyle{D}}}
\newcommand{\ssT}{{\scriptscriptstyle{T}}}

\newcommand{\ssW}{{\scriptscriptstyle{\rm W}}}

\newcommand{\ssM}{{\scriptscriptstyle{M}}}
\newcommand{\ssN}{{\scriptscriptstyle{N}}}

\long\def\symbolfootnote[#1]#2{\begingroup%
\def\thefootnote{\fnsymbol{footnote}}\footnote[#1]{#2}\endgroup}

\begin{document}


\begin{center}

{\Large \bf   
Weyl Semimetal and Topological Numbers}

\vspace{2cm}

\ Mahmut Elbistan

\vspace{5mm}

{\em {\it Institute of Modern Physics, Chinese Academy of Sciences, Lanzhou, China }}\footnote{{\it E-mail address:} elbistan@impcas.ac.cn }

\end{center}

\vspace{3cm}

Generalized Dirac monopoles in momentum space are constructed in even $d+1$ dimensions from the  Weyl Hamiltonian in terms of Green's functions. In $3+1$ dimensions, the (unit) charge of the monopole is equal to both the winding number and the Chern number, expressed as the integral of the Berry curvature.  Based on the equivalence of the Chern and winding numbers, a chirally coupled and Lorentz invariant field theory action is studied for the Weyl semimetal phase. At the one loop order, the effective action yields both the chiral magnetic effect and the anomalous Hall effect. The Chern number appears as a coefficient in the conductivity, thus emphasizes the role of topology. The anomalous contribution of chiral fermions to transport phenomena is reflected as the gauge anomaly with the Pfaffian invariant $(\bm{E}\cdot\bm{B})$.
Relevance of monopoles and Chern numbers for the semiclassical chiral kinetic theory is also discussed.

\vspace{11cm}

\pagebreak



\section{Introduction}
 
Recently, a fascinating condensed matter realization of chiral fermions has been proposed. This new topological phase is called  the Weyl semimetal \cite{wtvs, buba, ylr}. Its band structure has linear crossings at the so-called Weyl points where the system is effectively represented by the $2\times 2$ Weyl Hamiltonian. It is possible to derive a Berry potential \cite{berry} by considering the eigenstates of the Weyl Hamiltonian. The nontrivial topological properties of this novel phase \cite{wtvs, ylr, hq} is reflected by the flux of the Berry potential which is itself written as a Chern number (see \cite{dsPRD, ds} as a recent application to chiral fermions).

There are  both theoretical \cite{wtvs, buba, ylr, corzub, zubkov, zb, gs, g, jk, gms, liu} and experimental \cite{huangex, xuex, luex, wengex, lvex, zhangex} studies on Weyl semimetals. In their specific model Zyuzin and Burkov \cite{zb} showed that the topological transport properties of the Weyl semimetal, e.g., the chiral magnetic effect (CME) (see \cite{khar} and the references therein) and the anomalous Hall effect (AHE) \cite{hal}, are related to the chiral anomaly \cite{adler, bj, be}. 

Field theory models of Weyl semimetal \cite{zb, gs, g, gms} are constructed by means of Dirac spinors coupled to a Lorentz symmetry breaking $b_\mu$ vector \cite{colkos, jackos}. Axial coupling $\gamma^5 b_\mu$ is necessary to prevent the Weyl nodes to annihilate each other. However, these studies did not refer to the Chern number and its function, therefore the role of topology is not explicit. \footnote{We note that in \cite{gs} a numerical study of conductivity based on a lattice model is done by means of Berry curvature of two bands.}

In this study we first investigate the relations between topological numbers associated with the $3+1$ dimensional Weyl Hamiltonian. We construct a Dirac monopole of unit charge in momentum space in terms of the Green's function of the Weyl Hamiltonian.  On the $2$-dimensional boundary, the charge itself can be expressed as the Chern character of the Berry potential. It is also equal to the winding number of the Green's function.

Then, focusing on a single isolated node, we consider a chirally coupled, Lorentz invariant field theory model for the type-I Weyl semimetal where the dispersion relation and quasiparticles respect the emergent Lorentz symmetry \cite{horava, vz, v}. By integrating out the fermionic degrees of freedom via Feynman path integration, we relate the winding number (Chern number) to the Weyl semimetal explicitly and obtain an effective action. Akin to the case of quantum Hall effect \cite{tknn, ass}, the conductivity is written in terms of the topological Chern number. The resulting electromagnetic current has anomalous transport properties, namely CME and AHE. However, as opposed to the other field theory models, our quantized system exhibits the gauge anomaly \cite{agg1, agg2}. We will also discuss similarities and differences with other field theories of the Weyl semimetal \cite{zb, gs, g, jk, gms}. 

Chiral fermions are also used in semiclassical chiral kinetic theories \cite{soy, sy} where the chiral anomaly, a purely quantum mechanical phenomenon, can be realized. 
The generalization of the semiclassical chiral kinetic theory to higher dimensions has been achieved \cite{ds} using the orbit method.  Moreover in \cite{dsPRD, ds} the classical version of covariant, non-Abelian gauge anomalies are also derived. The relation between CME and the chiral anomaly in $3+1$ dimensions has already been found \cite{soy, sy}. A general formulation of both the semiclassical chiral anomaly and the semiclassical CME in $d+1$ dimensions is given in \cite{de}. We briefly argue in favor of the relevance of the topological numbers associated with the Weyl Hamiltonian within these semiclassical models. We also generalize the Dirac monopole for higher dimensional Weyl Hamiltonians.

The manuscript is organized as follows. In Section 2 we give a general presentation of the Berry potential and the Berry curvature associated with the Weyl Hamiltonian. In Section 3 we construct the Dirac monopole of unit charge in $3+1$ dimensions and point out its connection with the Chern and the winding numbers.  In Section 4 we study a low energy field theory Lagrangian for the Weyl semimetal and calculate the effective action by quantizing it.  In Section 5 we generalize the monopole construction to higher dimensions. In the last section we summarize and discuss our results. 

Henceforth, we use $\hbar=c=1$  throughout the manuscript.


\section{Weyl Hamiltonian and Berry Gauge Field }

In any even $d+1$ dimensional spacetime the one-particle, positive chirality Weyl Hamiltonian
\be
\label{hwg}
{\cal{H}}_\ssW=\bm{\Sigma}\cdot\bm{p},
\ee
is expressed in terms of the $d$ dimensional momentum vector $\bm{p}$ and the $2^{\frac{d-1}{2}}\times 2^{\frac{d-1}{2}}$ dimensional $\bm{\Sigma}$ matrices satisfying the relation $\{\Sigma_{\ssM},\Sigma_{\ssN}\}=2\delta_{\ssM\ssN}$ where $M, N=1,..,d$. The Weyl Hamiltonian (\ref{hwg}) is derived from the massless Dirac Hamiltonian,
$$
{\cal{H}}_{\scriptstyle{D}}=\bm{\alpha}\cdot\bm{p}, \quad \{\alpha_{\ssM},\alpha_{\ssN}\}=2\delta_{\ssM\ssN},
$$
which is block diagonal in the chiral representation of the $2^{\frac{d+1}{2}}\times 2^{\frac{d+1}{2}}$  dimensional $\bm{\alpha}$ matrices. The Weyl equation is solved for eigenvectors $|\psi_{\pm}^{(\alpha)}(\bm{p})\rangle$ as,
\be
\label{weyleq}
{\cal{H}}_\ssW|\psi_{\pm}^{(\alpha)}(\bm{p})\rangle=\pm |\bm{p}||\psi_{\pm}^{(\alpha)}(\bm{p})\rangle,
\ee
where $\alpha, \beta...=1,..,\frac{d-1}{2}$ indicating the $\frac{d-1}{2}$ fold degeneracy of the each eigenvalue $\pm |\bm{p}|$. 

One can find a unitary matrix $U$ which diagonalizes (\ref{hwg}) as
\be
\label{diag}
U{\cal{H}}_{\ssW} U^\dagger={\rm{diag}}(|\bm{p}|,-|\bm{p}|)=|\bm{p}|({\cal{I}}^+-{\cal{I}}^-),
\ee
where ${\cal{I}}^+$ and ${\cal{I}}^-$ are $2^{\frac{d-1}{2}}\times 2^{\frac{d-1}{2}}$ dimensional matrices projecting onto the positive and negative energy subspaces, respectively:
$$
\label{im}
{\cal{I}}^+=\begin{pmatrix} 1 & 0 \\ 0 & 0 \end{pmatrix}, \quad  {\cal{I}}^-=\begin{pmatrix} 0 & 0 \\ 0 & 1 \end{pmatrix}.
$$

The Berry connection ${\cal{\bm{A}}}$ is introduced by means of $U$ and ${\cal{I}}^+$ as
\be
\label{dbgu}
{\cal{\bm{A}}}=i{\cal{I}}^+U\partial_{\bm{p}}U^\dagger {\cal{I}}^+.
\ee  
${\cal{\bm{A}}}$ is Abelian in $3+1$ dimensional spacetime however, because of the $\frac{d-1}{2}$ fold degeneracy, it will be  non-Abelian in higher dimensions. Equivalently, (\ref{dbgu}) can also be written in terms of the positive energy solutions (\ref{weyleq}):
$$
{\cal{A}}^{\alpha\beta}_\ssM=i\langle\psi^{(\alpha)}_+|\partial_{p_{\ssM}}|\psi^{(\beta)}_+\rangle.
$$
The related Berry field strength is given as
\be
\label{dbfs}
{\cal{G}}^{\alpha\beta}_{\ssM\ssN}=\partial_\ssM{\cal{A}}^{\alpha\beta}_\ssN-\partial_\ssN{\cal{A}}^{\alpha\beta}_\ssM-i[{\cal{A}}_\ssM,{\cal{A}}_\ssN]^{\alpha\beta},
\ee
where the shorthand notation $\partial_\ssM=\partial_{p_\ssM}\equiv \frac{\partial}{\partial p^M}$ is used. Here, we consider only the positive energy eigenspace and exclude the level crossing point $|\bm{p}|=0$, the origin of the momentum space which is singular.

In order to acquire the Green's functions, we invert the relation (\ref{diag}) 
\be
\label{dpog}
{\cal{H}}_\ssW=|\bm{p}|(P^+-P^-),\quad P^{\pm}=U^{\dagger}{\cal{I}}^{\pm}U,
\ee
where $P^{\pm}=|\psi^{\pm}\rangle\langle\psi^{\pm}|$ are projection operators:
\be
\label{ppog}
P^++P^-=1,\quad P^{\pm}P^{\mp}=0,\quad P^{\pm}P^{\pm}=P^{\pm}.
\ee
They can also be expressed by means of ${\cal{H}}_\ssW$:
\be
\label{ph}
P^\pm=\frac{1}{2}(1\pm\frac{{\cal{H}}_\ssW}{|\bm{p}|}).
\ee
Now, we can write the Green's function $G(w,\bm{p})$ and its inverse $G^{-1}(w,\bm{p})$ for the Weyl Hamiltonian as
\be
\label{dgfg}
G(w,\bm{p})=\frac{P^+}{w-|\bm{p}|+i\epsilon}+\frac{P^-}{w+|\bm{p}|+i\epsilon}, \quad G^{-1}(w,\bm{p})=w-|\bm{p}|(P^+-P^-)+i\epsilon,
\ee
where $\epsilon$ is a small parameter.
Derivatives of $G^{-1}(w,\bm{p})$ with respect to $(w, \bm{p})$ are calculated to be
\be
\label{deriv}
\frac{\partial G^{-1}}{\partial p^0 }\equiv\frac{\partial G^{-1}}{\partial w }=1,\quad \frac{\partial G^{-1}}{\partial{p^\ssM }}=-\left( \frac{p_\ssM}{|\bm{p}|}(P^+-P^-)+|\bm{p}|\partial_{\ssM}(P^+-P^-)\right).
\ee


\section{$3+1$ Dimensional Weyl Hamiltonian and Topological Numbers}

In $3+1$ dimensional spacetime the left handed (positive chirality) Weyl Hamiltonian ${\cal{H}}^3_{\ssW}$ is expressed in terms of the Pauli spin matrices $\sigma_a$,
\be
\label{hw3}
{\cal{H}}^3_{\ssW}=\bm{\sigma}\cdot{\bm{p}}
=\begin{pmatrix} p_3 & p_1-ip_2\\ p_1+ip_2 & -p_3 \end{pmatrix}.
\ee
It can be diagonalized (\ref{diag}) with the unitary matrix
\be
\label{u3}
\displaystyle
U=\big(|\psi_+(\bm{p})\rangle\ |\psi_-(\bm{p})\rangle\big)^\dagger=\begin{pmatrix} N_+ & \frac{N_+(|\bm{p}|-p_3)}{p_1+ip_2} \\[10pt]
N_- & \frac{-N_-(|\bm{p}|+p_3)}{p_1+ip_2} \end{pmatrix}, \quad N_+={\scriptstyle{\sqrt{\left( \frac{|\bm{p}|+p_3}{2|\bm{p}|}\right)}}},
\quad N_-={\scriptstyle{\sqrt{\left( \frac{|\bm{p}|-p_3}{2|\bm{p}|}\right)}}}
\ee
which is constructed out of the  solutions of momentum space eigenvalue equation (\ref{weyleq}).

The construction of Dirac monopoles in terms of the Berry potential from the Weyl Hamiltonian (\ref{hwg}) was discussed and generalized to all even dimensions \cite{bfh} by using algebraic techniques. However, in order to make a connection with the Chern numbers, it is also useful to obtain these monopoles with the Green's function (\ref{dgfg}). Our approach has similarities with the one presented in \cite{ds}.

In $3+1$ dimensions we propose ${\cal{C}}_3$ in order to construct the associated monopole,
\begin{eqnarray}
\label{dwn3}
{\cal{C}}_3={\frac{1}{2\pi^2}\int{d^3p\ dw\ \epsilon^{abc}\Tr[(G\partial_0G^{-1})(G\partial_aG^{-1})(G\partial_bG^{-1})(G\partial_cG^{-1})]}},
\end{eqnarray}
where $\partial_0=\frac{\partial}{\partial p^0}$, $\partial_a=\frac{\partial}{\partial p^a}$ and $a,b,c=1,2,3$. $\Tr$ denotes the trace over the spin indices.

In order to calculate (\ref{dwn3}), we employ (\ref{ppog}), (\ref{dgfg}) and (\ref{deriv}). It is straightforward to observe that the quadratic and the cubic terms in $p_a$ vanish due to the antisymmetry of the Levi-Civita tensor. A careful investigation shows that terms linear in $p_a$ also do not  contribute after integration on  $w$. We find
\be
\label{cs3}
{\cal{C}}_3=-\frac{i}{2\pi}\int{d^3p\ \epsilon^{abc}\Tr[\partial_aP^+\partial_bP^+\partial_cP^+]}.
\ee
As $P^+$ is a $2\times 2$ matrix, (\ref{cs3}) does not vanish. However, the integrand is a total derivative,
$$
\int{d^3p\ \epsilon^{abc}\Tr[\partial_aP^+\partial_bP^+\partial_cP^+]}=\int{d^3p\ \bm{\nabla}\cdot \bm{K}_3}, 
$$
where we introduced $
K_3^a=\epsilon^{abc} \Tr[P^+\partial_bP^+\partial_cP^+]. 
$
Making use of (\ref{ph}) provides $K_3^a$ with a simple form and we calculate it as 
\begin{eqnarray}
\label{cp3}
K_3^a=\frac{1}{(2|\bm{p}|)^3}\epsilon^{abc}\Tr[{\cal{H}}^3_\ssW(\partial_b{\cal{H}}^3_\ssW)(\partial_c{\cal{H}}^3_\ssW)]
=\frac{ip^a}{2|\bm{p}|^3}.
\end{eqnarray}   
We conclude that (\ref{dwn3}) is the divergence of the field of a monopole $ \bm{b}_3=\frac{\bm{p}}{2|\bm{p}|^3}$ located at the center of the momentum space $|\bm{p}|=0$. We observe that ${\cal{C}}_3$ is equal to the unit charge of this monopole: 
\be
\label{wnmf3}
{\cal{C}}_3=\frac{1}{2\pi}\int{d^3p\ \bm{\nabla}\cdot\bm{b}_3}=1, \quad  \bm{\nabla}\cdot\bm{b}_3=2\pi\delta^3(|\bm{p}|)
\ee
This results seems to be in contradiction with the fact that the singular point $|\bm{p}|=0$ does not belong to our manifold.  However, (\ref{wnmf3}) should be thought as the effect of  non-trivial, quantum mechanical Berry's flux towards the positive energy eigenspace \cite{sy, dsPRD, ds}.\footnote{In fact, rigorously one should consider a very small region, say a sphere with radius $R$ around the singular point and convert our Eq. (16) into a surface integral.  Then, we calculate the related flux by taking the limit $R\to 0$.} Indeed, $|\bm{p}|=0$ is the point where the classical description of the chiral kinetic theories breaks down and why quantum anomaly occurs within those theories. 

Yet (\ref{cp3}) deserves a closer look. (\ref{dpog}) enables us to express $K_3^a$ by means of the matrix $U$ (\ref{u3}) as
$$
K_3^a=\epsilon^{abc}\Tr[{\cal{I}}^+(\partial_bU)(\partial_cU^\dagger){\cal{I}}^+].
$$
With the help of (\ref{dbgu}) and (\ref{dbfs}) we write the Abelian Berry curvature as
$$
\label{bfg3sym}
{\cal{G}}_{ab}=\partial_a{\cal{A}}_b-\partial_b{\cal{A}}_a=i{\cal{I}}^+\big((\partial_aU)(\partial_bU^\dagger)-(\partial_bU)(\partial_aU^\dagger)\big){\cal{I}}^+,
$$
so that
$
K_3^a=\frac{1}{2i}\epsilon^{abc}{\cal{G}}_{bc}.
$
This reveals the relation between ${\cal{C}}_3$ (\ref{dwn3}) and the Berry field strength:
$$
{\cal{C}}_3=-\frac{1}{4\pi}\int{d^3p\ \epsilon^{abc}\partial_a{\cal{G}}_{bc}}.
$$
We call our $3-$dimensional momentum space outside $|\bm{p}|=0$ as $\sigma_3$ whose boundary is homeomorphic to $S^2$.
We may express the flux of the monopole on the unit sphere $S^2$ as
\be
\label{bcn3}
{\cal{C}}_3=-\frac{1}{4\pi}\int_{\sigma_3}{d^3p\ \epsilon^{abc}\partial_a{\cal{G}}_{bc}}=-\frac{1}{4\pi}\int_{S^2}{d^2 p\ \epsilon^{\rm{b}\rm{c}}{\cal{G}}_{\rm{b}\rm{c}}},
\ee
where $\rm{b},\rm{c}$ represent the polar and the azimuthal angles $\theta , \phi$ respectively and ${\cal{G}}_{\theta\phi}=\frac{\sin\theta}{2}$. We recognize (\ref{bcn3}) as the minus the first topological Chern number.
The integral of the Berry curvature  over the compact manifold $S^2$ (Chern character) results in $-1$. \footnote{The calculation of (\ref{dwn3}) can be performed via the negative energy eigenspace as well. Then, one defines the Berry gauge field in terms of the negative energy solution $|\psi_-(p)\rangle$ which yields ${\cal{A}}^a=\frac{\epsilon^{ab3}p_b}{2|\bm{p}|(|\bm{p}|-p_3)}$. The value of (\ref{wnmf3}) and (\ref{bcn3}) will not alter under these changes as it is expected. However, calculation of the first Chern number in the negative eigenspace is $+1$. Thus, unlike ${\cal{C}}_3$, the Chern number is sensitive to the projection.} 

The expression for the Chern number (\ref{bcn3}) corresponds to the equations $\#$ $(B.14)$ and $\#$ $(B.15)$ of \cite{ds} in the $3+1$ dimensional case. This is not a surprise since our (\ref{cp3}) is of the same form as the Chern number construction of \cite{ds}, namely their equation $\#$ $(B.3)$. An expression similar to ${\cal{C}}_3$ is also used in \cite{zubkov} as a topological invariant. 

For further reasons, we define a 1-form gauge field ${\cal{A}}_a dp^a$
where ${\cal{A}}_a$ is the Berry potential (\ref{dbgu}). 
It is computed either in terms of the positive energy solution $|\psi_+\rangle$ or in terms of $U$
\be
\label{bgf3}
{\cal{A}}^a=\frac{\epsilon^{ab3}p_b}{2|\bm{p}|(|\bm{p}|+p_3)}.
\ee

We note that (\ref{bgf3}) is endowed with a Dirac string along the negative $p_3$-axis.  Gauge group of ${\cal{B}}_3$ is $U(1)$ and (\ref{bcn3}) is known to be the winding number of the principal bundle $P(S^2, U(1))$.

Monopole charge for ${\cal{H}}^3_{\ssW}$ is also expressed as an integral over $S^2$ in momentum space \cite{v}, 
\be
\label{wind3}
{\cal{C}}_3=\frac{\epsilon^{abc}}{8\pi}\int_{S^2} dn_a\ \hat{\bm{p}}\cdot\big(\frac{\partial \hat{\bm{p}}}{\partial p^b}\times\frac{\partial \hat{\bm{p}}}{\partial p^c}   \big) =\frac{1}{2\pi}\int_{S^2}d\bm{n}\cdot\bm{b}_3
\ee 
which is a topological invariant where $S^2$ encloses the level crossing point $|\bm{p}|=0$.

Remarkably, it can be written \cite{gv, v, vz, v2, v3, v4} as the topological winding number of the Green's function (\ref{dgfg}):
\be
\label{dwn3alt}
{\cal{C}}_3=-\frac{\textcolor{magenta}{}i}{24\pi^2}\int_\sigma{dS_\lambda}\ \epsilon^{\mu\nu\rho\lambda}\Tr[(G\partial_{\mu}G^{-1})(G\partial_{\nu}G^{-1})(G\partial_{\rho}G^{-1}).
\ee
Here $\sigma$ is a 3-dimensional surface closed around the Fermi point $|\bm{p}|=0$ where $G(w,\bm{p})$ is continuous and differentiable. The  integral (\ref{dwn3alt}) results in the chirality of the particle which equals to $1$ for our Hamiltonian (\ref{hw3}). 

We conclude that ${\cal{C}}_3$ (\ref{dwn3}) relates the topological Chern (\ref{bcn3}) and winding numbers (\ref{dwn3alt}) as it yields a momentum space monopole whose charge equals to its chirality. Indeed, choosing the opposite chirality Hamiltonian $-\bm{\sigma}\cdot{\bm{p}}$ we would obtain ${\cal{C}}_3=-1$. 

We emphasize that (\ref{dwn3}) is convenient to compute and to generalize to higher dimensions (see (\ref{cdk})). In the Appendix, we show explicitly that similar results hold in $5+1$ dimensions.


\section{Chiral Field Theory Model for Weyl Semimetals}

The low energy effective Hamiltonian of the Weyl semimetal is the $3+1$ dimensional Weyl Hamiltonian (\ref{hw3}) and the Weyl points behave like Dirac monopoles in momentum space (\ref{cp3},\ref{wnmf3}). In \cite{wtvs, ylr, hq} it was noted that the topological nontriviality of the Weyl semimetallic phase is related to the Chern number (\ref{bcn3}) which is built by means of the Berry curvature. 
Based on the equivalence of (\ref{bcn3}) and (\ref{dwn3alt}) through (\ref{dwn3}), we deduce that the topological properties of the Weyl semimetal phase is directly connected to the winding number of the Weyl fermion propagator. 

In \cite{qhz, dey} the relation between the winding number of the massive fermion propagator and the Chern number for topological insulators was investigated. There, the winding number happens to be the coefficient of the effective action which is derived by integrating out the fermionic degrees of freedom \cite{gjk}. We will follow a similar path here and search for a field theoretical description for the Weyl semimetal augmented with topological arguments.


\subsection{The Classical Action and Its Properties}

Since we focus on chiral fermions with positive chirality (only one Weyl node), we consider a Lagrangian of left handed Weyl spinors $\psi_L$ having only 2 non-zero components,
\be
\label{weyll}
{\cal{L}}_\ssW=\bar{\psi}_L(i\slashed{\partial}-e\slashed{A}-\slashed{B})\psi_L,\quad \psi_L(\bm{x}, t)=P_L\psi(\bm{x}, t),
\ee
where $\slashed{\partial}=\gamma^\mu \partial_\mu$ and the metric is $\rm{diag}(+, -,-,-)$.  $\psi(\bm{x}, t)$ is the Dirac spinor and $P_L$ is the operator projecting onto the positive chirality subspace. $A_\mu(\bm{x}, t)$ and $B_\mu (\bm{x}, t)$ are electromagnetic and auxiliary gauge fields, respectively. 

The idea of the chiral Lagrangian (\ref{weyll}) is in agreement with the results of \cite{horava} where the low energy excitations of the $3+1$ dimensional microscopic Fermi liquid system is considered. In the vicinity of the Fermi point, these excitations can be effectively written by means of emergent, coarse grained, gapless spinors of $SO(3,1)$ \cite{horava}. 

A similar Lagrangian (together with emergent gravity) is constructed explicitly \cite{vz} as an emergent theory of 2-component Weyl fermions where the other $U(1)$ gauge field $B_\mu$ originates from the interaction of the original Dirac fermions. There, the topological stability of the Fermi point is expressed by (\ref{wind3}). (\ref{weyll}) is also similar to the positive chirality part of the Weyl semimetal model presented in \cite{corzub} where $B_\mu$ field emerges from elastic deformations. 

The electromagnetic current $j^\mu_A=e\bar{\psi}_L\gamma^\mu\psi_L$ in (\ref{weyll})  is conserved, $\partial_\mu j^\mu_A=0$. There is another conserved current $j^\mu_B$ which is coupled to $B_\mu$ field. As a result, (\ref{weyll}) has the $U_A(1)\times U_B(1)$ gauge symmetry.

In order to obtain the effective action $S_{eff}[A(x), B(x)]$, one may naively  try to integrate out the spinors in (\ref{weyll}). However, the Weyl operator $(i\slashed{\partial}-e\slashed{A}-\slashed{B})P_L$ maps a left handed spinor to a right handed one and therefore, the fermionic determinant is not well-defined. 
As noted in \cite{agg1, agg2}, to overcome this mathematical difficulty, we add free right handed fermions to (\ref{weyll}) as
\be
\label{weyllplus}
{\cal{L}}'_\ssW=\bar{\psi}(i\slashed{\partial}-e\slashed{A}P_L-\slashed{B}P_L)\psi.
\ee

As a gauge theory ${\cal{L}}'_\ssW$ is equivalent to ${\cal{L}}_\ssW$, e.g.  it yields the same Feynman diagrams. They are both chiral theories and parity breaking. The only difference between them is that, in ${\cal{L}}'_\ssW$ we deal with the Dirac propagator instead of a Weyl propagator. The advantage of ${\cal{L}}'_\ssW$ is in its mathematical correctness to calculate the fermionic determinant. We point out that although we do not have an underlying microscopic model, the existence of both left and right handed fermions in (\ref{weyllplus}) is consistent with the Nielsen-Ninomiya theorem \cite{nn}. 

${\cal{L}}'_\ssW$ is endowed with $U^L_A(1)\times U^L_B(1)$ gauge symmetry
\begin{eqnarray*}
U^L_A(1)&:& \psi\to e^{i\alpha P_L}\psi,\quad \bar{\psi}\to\bar{\psi}e^{-i\alpha P_R},\quad A_\mu\to A_\mu-\frac{1}{e}\partial_\mu\alpha, \\
U^L_B(1)&:& \psi\to e^{i\beta P_L}\psi,\quad \bar{\psi}\to\bar{\psi}e^{-i\beta P_R},\quad B_\mu\to B_\mu-\partial_\mu\beta,
\end{eqnarray*} 
where only the left handed fermions transform. 

Considering only the left handed node in our model (\ref{weyll}) (or (\ref{weyllplus})) is plausible since left and right handed nodes are well separated in momentum space. We emphasize that the usage of ${\cal{L}}'_\ssW$ should not be understood as if only the left handed fermions interact with gauge fields in Weyl semimetals. Indeed, one can consider the negative chirality node, $-\bm{\sigma}\cdot\bm{p}$, as well. Then the right handed particles would seem to be interacting with the gauge fields while the left handed ones are free. 


\subsection{Path Integral and Anomalous Currents}

Integration over the spinor fields 
$$
\frac{\int {\cal{D}}\bar{\psi}{\cal{D}}\psi\  e^{i\int d^4x \bar{\psi}(i\slashed{\partial}-e\slashed{A}P_L-\slashed{B}P_L)\psi}  }{\int {\cal{D}}\bar{\psi}{\cal{D}}\psi\  e^{i\int d^4x \bar{\psi}i\slashed{\partial}\psi}}=e^{iS_{eff}[A, B]},
$$     
yields the following effective action:
\begin{eqnarray}
\label{fdet}
S_{eff}[A, B]=-i\ln\Big(\det\big[ 1-(e\slashed{A}+\slashed{B})P_LG_\ssD\big]\Big)
=-i\Tr\Big[\ln\big(1-(e\slashed{A}+\slashed{B})P_LG_\ssD\big) \Big].
\end{eqnarray}

Apart from the tree level term, expansion of (\ref{fdet}) yields many loop corrections. Vacuum polarization diagrams with two external legs are unrelated to the winding number (\ref{dwn3alt}) and will result with the half that of the Dirac theory, so we ignore them here. The next leading order terms are the triangle diagrams. Terms with three $A_\mu$ or $B_\mu$ external legs will vanish due to the antisymmetry of Levi-Civita tensor. Then, the physical response of the effective action $S_{eff}[A, B]$ to the external electromagnetic fields will come from the triangle diagram of two $A_\mu$ and one $B_\mu$ fields. The other triangle diagrams with two external $B_\mu$ field legs will be vanishing with our substitution $B_\mu(x)\equiv\partial_\mu\theta(x)$ (see Section 4.3). 

Thus, we will focus only on the $e^2$ terms with $3$ external legs consisting of two $A_\mu$ and one $B_\mu$ fields as
\begin{eqnarray*}
S_{eff}[A, B]=\frac{e^2(-i)^3}{3}\int d^4x_1 d^4x_2 d^4x_3\    
\Tr\Big[\slashed{A}(x_1) G_\ssD (x_2-x_1)\slashed{A}(x_2) G_\ssD (x_3-x_2)\slashed{B}(x_3) G_\ssD (x_1-x_3)P_R&\\
+\slashed{A}(x_1) G_\ssD (x_2-x_1)\slashed{B}(x_2) G_\ssD (x_3-x_2)\slashed{A}(x_3)G_\ssD (x_1-x_3)P_R&\\
+\slashed{B}(x_1) G_\ssD (x_2-x_1)\slashed{A}(x_2) G_\ssD (x_3-x_2)\slashed{A}(x_3) G_\ssD (x_1-x_3)P_R&
\Big].
\end{eqnarray*}
Here $G_\ssD(x-y)$ is the Dirac propagator,
$$
G_\ssD(x-y)=\int{\frac{d^4p}{(2\pi)^4}\frac{ie^{-ip\cdot(x-y)}}{\slashed{p}+i\epsilon}}.
$$
We express it and its inverse in momentum space as
\be
\label{gdirac}
G_\ssD(p)=\frac{i}{\slashed{p}+i\epsilon}=\frac{i\slashed{p}}{p^2+i\epsilon},\quad G_\ssD^{-1}(p)=-i(\slashed{p}+i\epsilon),
\ee
where $\epsilon$ is a small parameter and $p^2=p_\mu p^\mu$. We also perform the Fourier transformation of the external gauge fields as
$$
A_\mu(x)=\int{\frac{d^4k}{(2\pi)^4}e^{-ik\cdot x}A_\mu(k)}.
$$
We use the chiral representation of the gamma matrices,
$$
\gamma^0=
\begin{pmatrix}
0 & -1\\
-1 & 0
\end{pmatrix},\quad \gamma^i=\begin{pmatrix}
0 & \sigma^i\\
-\sigma^i & 0
\end{pmatrix},\quad \gamma^5=\begin{pmatrix}
1 & 0\\
0 & -1
\end{pmatrix},
$$ 
where $\gamma^5=i\gamma^0\gamma^1\gamma^2\gamma^3$.
Chirality projection operators $P_L$ and $P_R$ 
$$
\label{prpl}
P_L=\frac{1+\gamma^5}{2},\quad P_R=\frac{1-\gamma^5}{2},
$$  
satisfy the following relations
$$
\label{pcpo}
P_R+P_L=1,\quad P_L^2=P_L,\quad P_R^2=P_R,\quad P_LP_R=P_RP_L=0.
$$

Conservation of momenta in each vertex results in the following expression for the effective action 
$$
S_{eff}[A, B]=S_{eff}^1 +S_{eff}^2+S_{eff}^3
$$ 
where
\begin{subequations}
\label{effactionps}
\begin{align}
S_{eff}^1[A, B]=\frac{e^2}{3}\int{\frac{d^4k_1}{(2\pi)^4}}{\frac{d^4k_2}{(2\pi)^4}}A_\mu(k_1)A_\nu(k_2)B_\rho(-k_1-k_2)\Pi^{\mu\nu\rho}(k_1, k_2),   \\[8pt]
S_{eff}^2[A, B]=\frac{e^2}{3}\int{\frac{d^4k_1}{(2\pi)^4}}{\frac{d^4k_2}{(2\pi)^4}}A_\mu(k_1)B_\nu(k_2)A_\rho(-k_1-k_2)
\Pi^{\mu\nu\rho}(k_1, k_2),   \\[8pt]
S_{eff}^3[A, B]=\frac{e^2}{3}\int{\frac{d^4k_1}{(2\pi)^4}}{\frac{d^4k_2}{(2\pi)^4}}B_\mu(k_1)A_\nu(k_2)A_\rho(-k_1-k_2)\Pi^{\mu\nu\rho}(k_1, k_2).   
\end{align}
\end{subequations}
Here $k_1, k_2$ are the momenta of the external legs. $\Pi^{\mu\nu\rho}(k_1, k_2)$ is the expression for the triangle graph of the left handed fermions, 
\begin{eqnarray}
\label{righttriangle}
\Pi^{\mu\nu\rho}(k_1, k_2)&=&\int{\frac{d^4p}{(2\pi)^4}\Tr\Big[(-i\gamma^\mu) G_\ssD(p) (-i\gamma^\nu) G_\ssD(p+k_2) (-i\gamma^\rho) G_\ssD(p-k_1)\big(\frac{1-\gamma^5}{2}\big)\Big] }\nonumber\\
&=&\int{\frac{d^4p}{(2\pi)^4}\frac{\Tr\Big[\gamma^\mu \slashed{p}\gamma^\nu (\slashed{p}+\slashed{k_2})\gamma^\rho(\slashed{p}-\slashed{k_1})\frac{(1-\gamma^5)}{2}\Big]}{p^2(p+k_2)^2(p-k_1)^2}  },
\end{eqnarray}
where $p$ is the loop momentum and $i\epsilon$ factors are omitted. Only the part with the factor $-\frac{\gamma^5}{2}$ will yield non-zero contribution \cite{Srednicki}. 
The integral (\ref{righttriangle}) is linearly divergent and needs regularization. 

On the other hand we observe that the classically conserved currents $j^\mu_A$ and $j^\mu_B$ are no more conserved at the quantum level. In the case of $S_{eff}^1$ (\ref{effactionps}a) we calculated,
\be
\label{ncon1}
\partial_\mu \langle j^\mu_B(x)\rangle=0,\quad
\partial_\mu \langle j^\mu_A(x)\rangle= \frac{e^2}{24\pi^2}\epsilon^{\mu\nu\rho\lambda}\partial_\mu B_\nu \partial_\rho A_\lambda.
\ee
For the $S_{eff}^2+ S_{eff}^3$ (\ref{effactionps}b-c) we found,
\be
\label{ncon2}
\partial_\mu \langle j^\mu_B(x)\rangle= \frac{e^2}{24\pi^2}\epsilon^{\mu\nu\rho\lambda}\partial_\mu A_\nu \partial_\rho A_\lambda,\quad  \partial_\mu \langle j^\mu_A(x) \rangle= \frac{e^2}{24\pi^2}\epsilon^{\mu\nu\rho\lambda}\partial_\mu B_\nu \partial_\rho A_\lambda. 
\ee


\subsection{Winding Number and Effective Action}

$S_{eff}[A, B]$ should be compatible with (\ref{ncon1}) and (\ref{ncon2}). In order to calculate it we recall the winding number ${\cal{C}}_3$ (\ref{dwn3alt}) that is written in terms of $G(w,\bm{p})$ (\ref{dgfg}). This will also help us to incorporate topology in our calculation.

We start with decomposing $G_\ssD$ (\ref{gdirac}) into its chiral parts,
\begin{eqnarray*}
G_\ssD=P_LG_\ssD+P_RG_\ssD
\equiv G_\ssW^L+G_\ssW^R
\end{eqnarray*}
where $G_\ssW^{L,R}$ is the Weyl propagator corresponding to the left or right handed sector, respectively. As a $4\times 4$ matrix $G_\ssW^L$ can be written as
\be
\label{gwr}
G_\ssW^L=-\frac{i}{p^2}
\begin{pmatrix}
0 & p^0+\bm{p}\cdot\bm{\sigma}\\
0 & 0
\end{pmatrix},
\ee 
where we have omitted $i\epsilon$ factors.
We define
$$
\bar{\sigma}^\mu=(1,-\bm{\sigma}),\quad 
\sigma^\mu=(1,\bm{\sigma})
$$
to express a new $2\times 2$ matrix $[G_\ssW^L]$ which is the non-zero subblock of (\ref{gwr}),
\be
\label{gwrblock}
[G_\ssW^L]=\frac{-ip_\mu\bar{\sigma}^\mu}{p^2},\quad 
[G^L_\ssW]^{-1}=ip_\mu\sigma^\mu=i(p^0-\bm{p}\cdot\bm{\sigma}).
\ee   
This provides the relation between $G(w, \bm{p})$ introduced in the Hamiltonian formalism  and (\ref{gwrblock})
\be
\label{2g}
[G_\ssW^L]=-iG, \quad 
[G^L_\ssW]^{-1}=iG^{-1}, \quad 
[G^L_\ssW]
[G^L_\ssW]^{-1}=GG^{-1}=1,
\ee
where we have identified $w=p^0$. 
(\ref{2g}) is the main connection between topological arguments presented in the previous section and the field theory stated in (\ref{weyllplus}). 

Armed with these, our aim is to describe the winding number (\ref{dwn3alt}) in terms of the Dirac propagator $G_\ssD$. One may rewrite it with the help of the divergence theorem as
\begin{eqnarray*}
\label{dwn3volovik}
{\cal{C}}_3&=&-\frac{i}{24\pi^2}\int d^4p\  \epsilon^{\mu\nu\rho\lambda}\partial_\lambda\Tr\big[(G\partial_{\mu}G^{-1})(G\partial_{\nu}G^{-1})(G\partial_{\rho}G^{-1})\big].
\end{eqnarray*}
Using (\ref{2g}) we express the winding number ${\cal{C}}_3$  in terms $G_\ssD$ as
\begin{eqnarray}
\label{weyltodirac}
{\cal{C}}_3&=&-\frac{i}{24\pi^2}\int d^4p \epsilon^{\mu\nu\rho\lambda}\partial_\lambda\Tr\big[( [G^L_\ssW]\partial_{\mu}[G^L_\ssW]^{-1})([G^L_\ssW]\partial_{\nu}[G^L_\ssW]^{-1})([G^L_\ssW]\partial_{\rho}[G^L_\ssW]^{-1})\big]\nonumber\\
&=&-\frac{i}{24\pi^2}\int d^4p \epsilon^{\mu\nu\rho\lambda}\partial_\lambda\Tr\big[(P_L G_\ssD\partial_{\mu}G_\ssD^{-1})(P_LG_\ssD\partial_{\nu}G_\ssD^{-1})(P_LG_\ssD\partial_{\rho}G_\ssD^{-1})\big]\\
&=&-\frac{i}{24\pi^2}\int{ d^4p\ \epsilon^{\mu\nu\rho\lambda}\partial_{\lambda}\Tr\big[(\partial_{\mu}G_\ssD^{-1})
(G_\ssD\partial_{\nu}G_\ssD^{-1})
(G_\ssD\partial_{\rho}G_\ssD^{-1})
G_\ssD P_R\big]}.\nonumber
\end{eqnarray}

We are now in position to relate ${\cal{C}}_3$ to $\Pi^{\mu\nu\rho}$ (\ref{righttriangle}). In the weak field approximation, where the external momenta are vanishing, (\ref{weyltodirac}) can be written as
\begin{subequations}
\begin{align}
\label{ctop}
{\cal{C}}_3&=\frac{2i\pi^2}{3}\epsilon_{\mu\nu\rho\lambda}\frac{\partial}{\partial_{k_{1\lambda}}}\Pi^{\mu\nu\rho}(k_1,k_2)|_{k_{1,2}=0}\\
&=-\frac{2i\pi^2}{3}\epsilon_{\mu\nu\rho\lambda}\frac{\partial}{\partial_{k_{2\lambda}}}\Pi^{\mu\nu\rho}(k_1,k_2)|_{k_{1,2}=0}.
\end{align}
\end{subequations}
(\ref{ctop}-b) show that $\Pi^{\mu\nu\rho}(k_1, k_2)$ is a linear function of $k_1-k_2$. Indeed, we solve for $\Pi^{\mu\nu\rho}$ as
\be
\label{ptoc}
\Pi^{\mu\nu\rho}(k_1, k_2)=\frac{3i\ {\cal{C}}_3}{4!\ 2\pi^2}\epsilon^{\mu\nu\rho\lambda}(k_1-k_2)_{\lambda}.
\ee
Considering (\ref{ncon1}-\ref{ncon2}) and (\ref{ptoc}), we find  $S_{eff}^1=0$ and the effective action $S_{eff}=S_{eff}^2+S_{eff}^3$  results in 
\be
\label{pre-effective}
S_{eff}[A, B]=\frac{e^2 {\cal{C}}_3}{ 24\pi^2}\int d^4x \epsilon^{\mu\nu\rho\lambda} B_\mu A_\nu \partial_\rho A_\lambda.
\ee
(\ref{pre-effective}) looks odd since it is not gauge invariant under the action of $U_A(1)$ and $U_B(1)$, the symmetries of the classical theory (\ref{weyllplus}). However, we recall that the massless fermions end up in anomalies whenever they are quantized. 

In order to restore $U_A(1)$ electromagnetic gauge symmetry  we have to define $B_\mu$ as a pure gauge, $B_\mu(x)\equiv\partial_\mu\theta(x)$. Indeed, this definition is the only choice to be consistent with (\ref{ncon1}-\ref{ncon2}). Then, we have
\be
\label{boundary}
S_{eff}[A, B]=\frac{e^2 {\cal{C}}_3}{ 24\pi^2}\int d^4x \epsilon^{\mu\nu\rho\lambda}\partial_\mu(\theta A_\nu \partial_\rho A_\lambda)-\frac{e^2 {\cal{C}}_3}{ 24\pi^2}\int d^4x \epsilon^{\mu\nu\rho\lambda}\ \theta \partial_\mu A_\nu \partial_\rho A_\lambda.
\ee
Ignoring the total derivative term, we obtain the final form of the effective action:
\be
\label{effective}
S_{eff}[A, \theta]=-\frac{e^2 {\cal{C}}_3}{ 24\pi^2}\int d^4x \epsilon^{\mu\nu\rho\lambda}\ \theta \partial_\mu A_\nu \partial_\rho A_\lambda= -\frac{e^2 {\cal{C}}_3}{96 \pi^2}\int d^4x \epsilon^{\mu\nu\rho\lambda}\ \theta F_{\mu\nu}F_{\rho\lambda},
\ee
where $F_{\mu\nu}=\partial_\mu A_\nu-\partial_\nu A_\mu$ is the electromagnetic field tensor. 
(\ref{effective}) is gauge invariant with respect to $U_A(1)$ and the electromagnetic current is conserved, $\partial_\mu \langle j^\mu_A\rangle=0$. However, it is not invariant under the action of $U_B(1)$
$$
B_\mu\to B_\mu+\partial_\mu\lambda \quad \text{or}\quad \theta\to\theta+\lambda, 
$$
reflecting the anomalous contribution of chiral fermions. The corresponding nonconservation of the $B_\mu$ current is found as
\be
\label{gaugeanomaly}
\partial_\mu \langle j^\mu_B\rangle=\frac{e^2 {\cal{C}}_3}{96 \pi^2}\epsilon^{\mu\nu\rho\lambda}F_{\mu\nu}F_{\rho\lambda}=-\frac{e^2 {\cal{C}}_3}{12 \pi^2}(\bm{E}\cdot\bm{B}), 
\ee 
and this implies that we have a gauge anomaly for $U_B(1)$ symmetry. (\ref{gaugeanomaly}) is consistent with \cite{agg1, agg2, bardeen} (see also \cite{be}). It is also equal to (\ref{ncon2}) as it should be. We mention that similarly (consistent) gauge anomalies are also derived for chiral fermions in the (semi) classical context \cite{dsPRD, ds}.

At the classical level the currents $j_A(x)$ and $j_B(x)$ are both conserved. Nevertheless, at the quantum level, it is not possible to conserve them at the same time. We choose to conserve the electromagnetic current $\langle j^\mu_A\rangle$ because it is the physically relevant one. 

Although the effective action (\ref{effective}) is of the same form as the one obtained in \cite{zb}, the field theory (\ref{weyllplus}) is substantially different from the one there. In \cite{zb}, the low energy field theory was derived from a microscopic lattice model with the minimum number of two Weyl nodes consisting positive and negative chiralities. Therefore, they deal with a $4-$component Dirac spinor that includes Weyl spinors, both coupled to the electromagnetic gauge field. The Lorentz breaking vector $b_\mu$, with its axial coupling, separates the monopole and anti-monopole pair. $b_\mu$ also breaks parity and time reversal symmetries.

Instead of this particular realization, we deal with the generic model of \cite{vz} in order to explore the universal topological properties. Our parity non-invariant theory  (\ref{weyllplus}) contains only one Weyl node and it is already built by means of the chiral coupling of Weyl fermions to the gauge fields. As a result, instead of having chiral anomaly, we obtain a gauge anomaly with the topological term $(\bm{E}\cdot\bm{B})$. 

Ignoring the boundary term in (\ref{boundary}), we only focus on the bulk theory here. However, one should consider the effect of the boundary in a realistic condensed matter system. Due to our calculations electromagnetic gauge invariance dictates $B_\mu$ to be a pure gauge field. The related anomaly of the single positive chirality Weyl node (\ref{gaugeanomaly}) may be compensated with the help of a boundary theory.
Since Weyl nodes always come in pairs of opposite chirality \cite{nn}, the positive chirality Weyl node may be connected to the negative chirality one through the boundary. 
Thus, (\ref{gaugeanomaly}) may be interpreted as the manifestation of the surface states of the Weyl semimetal propagating between opposite chirality Weyl nodes, namely the Fermi arcs.


\subsection{Chiral Magnetic and Anomalous Hall Effects}

Now we discuss the expectation value of the electromagnetic current,
\be
\label{currentA}
\langle j^\mu_A\rangle\equiv\frac{\delta S_{eff}[A,\theta]}{\delta A_\mu}=-\frac{e^2 {\cal{C}}_3}{12 \pi^2} {\star F}^{\mu\nu}\partial_\nu \theta,
\ee
where ${\star F}^{\mu\nu}=\frac{1}{2}\epsilon^{\mu\nu\rho\sigma}F_{\rho\sigma}$ is the dual electromagnetic tensor with ${\star F}^{0i}=-B^i$ and ${\star F}^{ij}=\epsilon^{ijk}E^k$. 
We can decompose (\ref{currentA}) as
\begin{subequations}
\begin{align}
\langle  j^0_A (\bm{x}, t)\rangle\equiv n=\frac{e^2{\cal{C}}_3}{12\pi^2}\bm{\nabla}\theta\cdot\bm{B},\\
\label{CMEAHE}
\langle \bm{j}_A (\bm{x},{t})\rangle= -\frac{e^2{\cal{C}}_3}{12\pi^2}\big( (\bm{\nabla}\theta)\times\bm{E}+(\partial_t\theta)\bm{B}  \big),
\end{align}
\end{subequations}
where $n$ is the charge density and $\partial_t=\frac{\partial}{\partial t}$.
The current $\langle  \bm{j}_A\rangle$ has two terms which are particular to chiral fermions. The first term which is perpendicular to the electric field is called AHE. The charge density $n$ and the Hall conductivity $\sigma^{ij}_H$ are related as
\be
\label{Hconductivity}
\sigma^{ij}_H=\frac{e^2{\cal{C}}_3}{12\pi^2}\epsilon^{ijk}\partial_k\theta=\epsilon^{ijk}\frac{\partial n}{\partial B^k}.
\ee
The other part which is parallel to the magnetic field is the CME term with the magnetic conductivity $\sigma_M$:
\be
\label{Mconductivity}
\sigma_M=-\frac{e^2{\cal{C}}_3}{12\pi^2}\partial_t\theta.
\ee
We would like emphasize that (\ref{Mconductivity}) is insensitive to the temperature. 

Most importantly the nontrivial Chern number ${\cal{C}}_3$ (the winding number) contributes to both conductivities  (\ref{Hconductivity}, \ref{Mconductivity}) and highlights the role of the topology for the Weyl semimetal phase.  

We note that AHE and CME of chiral fermions were also derived within the (semi)classical chiral kinetic theories \cite{soy, sy, de, EM-HP-g0}.

We may also consider the contribution from the kinetic terms of the gauge fields to (\ref{weyllplus}). Since $B_\mu$ is a pure gauge field, it has no curvature. Then, (\ref{effective}) will modify the Maxwell's equations originating from the electromagnetic action $S_{EM}=-\frac{1}{4}\int d^4x F_{\mu\nu}F^{\mu\nu}$. The total action $S_{eff}+S_{EM}$ yields the modified equations,
\be
\label{tmeim}
\partial_\mu F^{\mu\nu}+\frac{e^2{\cal{C}}_3}{12\pi^2}(\partial_\mu\theta) {\star F}^{\nu\mu}=0,
\ee
which we write in the following explicit form:
\begin{subequations}
\label{tmeex}
\begin{align}
\bm{\nabla}\cdot\bm{E}-\frac{e^2{\cal{C}}_3}{12\pi^2}(\bm{\nabla}\theta)\cdot\bm{B}=0,\\
\bm{\nabla}\times\bm{B}-\partial_t \bm{E}+\frac{e^2{\cal{C}}_3}{12\pi^2}\big( (\partial_t\theta)\bm{B}+(\bm{\nabla}\theta)\times\bm{E}\big)=0.
\end{align} 
\end{subequations}
This modification leads to the topological magnetoelectric effect (TME) \cite{qhz}.
The definition of the bulk electric polarization $\bm{P}$, which is $\bm{j}_A=\partial_t\bm{P}$, implies
$$
\partial_t\bm{P}=-\frac{e^2{\cal{C}}_3}{12\pi^2}(\partial_t\theta)\bm{B}\implies \bm{P}=-\frac{e^2{\cal{C}}_3\theta}{12\pi^2}\bm{B}.
$$
We may also define the bulk magnetization $\bm{M}$ as $\bm{j}_A=\bm{\nabla}\times\bm{M}$ so that it is written in terms of the electric field: 
$$
\bm{\nabla}\times\bm{M}= -\frac{e^2{\cal{C}}_3}{12\pi^2}\bm{\nabla}\theta\times\bm{E}\implies \bm{M}=  -\frac{e^2{\cal{C}}_3\theta}{12\pi^2}\bm{E},
$$
where the electromagnetic fields are constant. 

We conclude that the Dirac monopole and thus the nontrivial Chern number ${\cal{C}}_3$ (\ref{wnmf3}) is responsible for the existence of TME (\ref{tmeim}, \ref{tmeex}) in our model. We note that in $3+1$ dimensional topological insulators, there exists a manifestation of TME with an image magnetic monopole in real space \cite{qlzz}.


\section{$d+1$ Dimensional Weyl Hamiltonian and Monopoles}

In this section, we generalize (\ref{dwn3}) to higher dimensions and search its relation to topological numbers.

In any even $d+1$ dimensional spacetime, each eigenvalue $(|\bm{p}|,-|\bm{p}|)$ of the Weyl Hamiltonian (\ref{hwg}) is $\frac{d-1}{2}$ fold degenerate and in principle the corresponding eigenstates 
$$
|\psi_+^{(1)}\rangle,...,|\psi_+^{({\scriptscriptstyle{{\frac{d-1}{2}}}})}\rangle, |\psi^{(1)}_-\rangle,...,|\psi_-^{({\scriptstyle{\frac{d-1}{2}}})}\rangle,
$$
can be computed. One can diagonalize the Weyl Hamiltonian with the unitary matrix $U,$
$$
U=\big(|\psi_+^{(1)}\rangle... |\psi_-^{({\scriptstyle{\frac{d-1}{2}}})}\rangle\big)^\dagger 
$$
and define the Berry connection (\ref{dbgu}) and its curvature (\ref{dbfs}).
With an appropriate normalization, we express the monopole charge (and the chirality) as \footnote{Actually, ${\cal{C}}_d$ should be multiplied with $-1$ for the $3+1$ dimensional case in order to cancel the minus factor which will appear in the $K_3^M$ (\ref{cpd}). This sign confusion is artificial in the sense that it is due to our choice of ${\cal{H}}^3_{\ssW}=\bm{\sigma}\cdot\bm{p}$ which is the conventional Weyl Hamiltonian used in the literature. If the Dirac matrices were constructed starting from the $1+1$ dimensions in the chiral basis, there would not be this sign ambiguity.}
\be
\label{cdk}
{\cal{C}}_{d}=\frac{i^{\frac{d+1}{2}}2^{\frac{3d-5}{2}}}{\pi (d-1)! {d-1 \choose \frac{d+1}{2}}{\rm{Vol}}(S^{d-1}) }\int{d^dp\ dw\ \epsilon^{MN..R}\Tr[(G\partial_{0}G^{-1})(G\partial_{M}G^{-1})...(G\partial_{R}G^{-1})]},
\ee 
where $M,N...=1,...d$.
Using (\ref{ppog}), (\ref{dgfg}) and (\ref{deriv}) one can perform the $w$ integration and obtain
\begin{eqnarray*}
{\cal{C}}_{d}&=&\frac{i^{\frac{d+1}{2}}2^{\frac{3d-5}{2}}}{\pi (d-1)! {d-1 \choose \frac{d+1}{2}}{\rm{Vol}}(S^{d-1}) }\int{d^dp\ dw\ \epsilon^{\ssM \ssN ... \ssR}\Tr[G^2(\partial_\ssM G^{-1})(G\partial_\ssN G^{-1})...(G\partial_\ssR G^{-1})]}\nonumber\\
&=&\frac{-2(-2i)^{\frac{d-1}{2}}}{(d-1)! {\rm{Vol}}(S^{d-1})}\int{d^dp\ \epsilon^{\ssM ...\ssR}\Tr[(\partial_\ssM P^+)...(\partial_\ssR P^+)]}\\
&=&\frac{-2(-2i)^{\frac{d-1}{2}}}{(d-1)! {\rm{Vol}}(S^{d-1})}\int d^dp\ {\bm{\nabla}\cdot\bm{K}_d}.\nonumber
\end{eqnarray*}
$\bm{K}_d$ is defined as
$
K_d^\ssM=\epsilon^{\ssM \ssN...\ssR}\Tr[ P^+(\partial_\ssN P^+)...(\partial_\ssR P^+)].
$
It leads to a Dirac monopole in $d$ dimensional momentum space:
\begin{eqnarray}
\label{cpd}
K_d^\ssM=\frac{1}{(2|\bm{p}|)^d}\epsilon^{\ssM \ssN...\ssR}\Tr[{\cal{H}}_\ssW(\partial_\ssN{\cal{H}}_\ssW)...(\partial_\ssR{\cal{H}}_\ssW)]
=-(d-1)! \big(\frac{i}{2}\big)^{\frac{d-1}{2}}\frac{p^\ssM}{2|\bm{p}|^d}.
\end{eqnarray} 
Thus, for all even spacetime dimensions ${\cal{C}}_{d}$ is associated with the monopole field $\bm{b}_d=\frac{\bm{p}}{2|\bm{p}|^d}$ and $\bm{\nabla}\cdot\bm{b}_d=\frac{{\rm{Vol}}(S^{d-1})}{2}\delta^d(|\bm{p}|)$. We calculate
\be
\label{wnmfd}
{\cal{C}}_{d}= \frac{2}{{\rm{Vol}}(S^{d-1})}\int{d^dp\ \bm{\nabla}\cdot\bm{b}_d}=1, 
\ee
and observe that ${\cal{C}}_d$ is equal to the unit charge of the Dirac monopole. As it is explained in $3+1$-d case, (\ref{wnmfd}) should be understood as the effect of non-trivial Berry's phase to the positive energy eigenspace.  

We note that, with a different point of view, a similar expression to (\ref{cpd}) is derived in \cite{ds}, namely their equation $\#$ $(B.12)$.

We can express (\ref{cdk}) in terms of the Berry curvature ${\cal{G}}^{\alpha\beta}_{\ssM\ssN}$ (\ref{dbfs}) and enclose the monopole with the compact surface $S^{d-1}$ 
\begin{eqnarray*}
{\cal{C}}_d&=&\frac{-2(-1)^{\frac{d-1}{2}}}{{\rm{Vol}}(S^{d-1})(d-1)!}\int_{\sigma_d}d^dp\ \epsilon^{\ssM \ssN\ssR...\ssS\ssT}\partial_\ssM \Tr [\overbrace{{\cal{G}}_{\ssN\ssR}...{\cal{G}}_{\ssS\ssT}}^{\text{(d-1)/2 times}}]\\
&=&\frac{2(-1)^{\frac{d+1}{2}}}{{\rm{Vol}}(S^{d-1})(d-1)!}\int_{S^{d-1}}d^{d-1}p\ \epsilon^{\rm{\ssN\ssR...\ssS\ssT}}\Tr[\overbrace{{\cal{G}}_{\rm\ssN\ssR}...{\cal{G}}_{\rm{\ssS\ssT}}}^{\text{(d-1)/2 times}}],
\end{eqnarray*} 
where $\sigma_d$ encloses the Fermi point $|\bm{p}|=0$.  $\rm{M,N...}$ represent the angular coordinates of $S^{d-1}$.
We observe that $\bm{b}_d$ can be expressed by means of the Berry curvature 
\be
\label{mfg}
b_d^\ssM=\frac{(-1)^{\frac{d+1}{2}}}{(d-1)!}\epsilon^{\ssM \ssN\ssR...\ssS\ssT} \Tr[\overbrace{{\cal{G}}_{\ssN\ssR}...{\cal{G}}_{\ssS\ssT}}^{\text{(d-1)/2 times}}]. 
\ee
\textcolor{red}{
}
(\ref{mfg}) is in accord with our previous result (\ref{bcn3}).  

Moreover, it demonstrates that the monopole which is responsible for the semiclassical version of the chiral anomaly within the chiral kinetic theory \cite{soy, sy, ds, de} is the same as the one in (\ref{wnmfd}). Since we show that ${\cal{C}}_d$ yields a monopole for any $d+1$ dimensional spacetime, we have proven the existence of the semiclassical chiral anomaly for all even dimensions implicitly. 

Let $d=2n+1$ where $n$ is a positive integer. The volume of $S^{d-1}$ is given in terms of $\Gamma$ function as $\frac{2\pi^{d/2}}{\Gamma(d/2)}$. We find
\begin{eqnarray*}
\label{bcnd}
{\cal{C}}_{2n+1}&=&\frac{\Gamma (n+\frac{1}{2})(-1)^{n+1}}{\pi^{n+\frac{1}{2}}\Gamma(2n+1)}\int_{S^{2n}}d^{2n}p\ \epsilon^{\rm{\ssN\ssR...\ssS\ssT}}\Tr[\overbrace{{\cal{G}}_{{\rm\ssN\ssR}}...{\cal{G}}_{{\rm\ssS\ssT}}}^{n\  times}]\nonumber \\
&=&\frac{(-1)^{n+1}}{(4\pi)^n n!}\int_{S^{2n}}d^{2n}p\ \epsilon^{\rm{\ssN\ssR...\ssS\ssT}}\Tr[\overbrace{{\cal{G}}_{\rm{\ssN\ssR}}...{\cal{G}}_{\rm{\ssS\ssT}}}^{n\  times}]=(-1)^{n+1}{\cal{N}}_n,
\end{eqnarray*}
and comparing it with (\ref{wnmfd}) we conclude that, 
\be
\label{nn}
{\cal{N}}_n=(-1)^{n+1},
\ee
for all $n$. Up to a minus factor, (\ref{nn}) is in accord with the spin Chern number computed in \cite{ds}, cf. equations $\#$ $(B.14)$ and $\#$ $(B.15)$ there.

Lastly, we investigate the gauge field structure of the $d$ dimensional Dirac monopole (\ref{wnmfd}). In \cite{n} the generalization of the Dirac monopoles to all dimensions is considered by means of the antisymmetric tensor gauge fields. Using differential forms, we observe
$$
\Tr[{\cal{G}}^n]=dQ_{\scriptscriptstyle{C}\ssS}^{2n-1},
$$
where 
$Q_{\scriptscriptstyle{C}\ssS}^{2n-1}$ is the $(2n-1)^{th}$ Chern-Simons form.  Therefore, we find the Abelian antisymmetric tensor gauge field ${\cal{B}}_{2n+1}$ of the monopole (\ref{mfg}) as
\be
\label{agfd}
{\cal{B}}_{2n+1}=Q_{\scriptscriptstyle{C}\ssS}^{2n-1}.
\ee


\section{Discussion}

We define the $d+1$ dimensional Berry potential and Berry curvature from the Weyl Hamiltonian. Both in $3+1$ and $5+1$ (see Appendix) dimensions, we explicitly express the Dirac monopole of unit charge in terms of Green's functions. The charge of the monopole is equal to the chirality of the Weyl Hamiltonian. We also show that the unit charge of the monopole can be written as the Chern number which is the integral of the Berry curvature. In $3+1$ dimensions, the monopole charge can also be expressed as the winding number of the fermion propagator.

Based on the arguments of \cite{horava, vz}, we propose a $3+1$ dimensional continuum field theory action for the Weyl semimetal. Since our chiral model (\ref{weyllplus}) is based on the positive chirality Weyl node, only the left handed fermions interact with the gauge fields. Therefore it is parity violating. 

Integration of the fermions yields an effective action which has the gauge anomaly  (\ref{gaugeanomaly}), proportional to Pfaffian invariant $\frac{1}{2}\epsilon^{\mu\nu\rho\lambda}F_{\mu\nu}F_{\rho\lambda}=-4\bm{E}\cdot\bm{B}$. As being equal to the winding number, Chern number happens to be the coefficient of the effective action (\ref{effective}). 
  
We have successfully obtained CME and AHE (\ref{CMEAHE}) which are particular anomalous transport properties of the Weyl semimetal. Topologically quantized Chern number also appears in the Hall conductivity  (\ref{Hconductivity}) and emphasizes the role of  topology for the Weyl semimetal. (\ref{effective}) also yields the topological magnetoelectric effect (\ref{tmeim}, \ref{tmeex}). Our model is Lorentz invariant, thus it may be a low energy, effective field theory candidate for the type-I Weyl semimetals.

We have to mention that there is a debate on the existence of CME in equilibrium (see \cite{changyang} and references therein). Although we derived CME in our model in the basis of topological arguments, we believe this issue deserves further investigation but it is beyond the scope of this work. 

The relation of the total derivative term (\ref{boundary}) and the gauge anomaly (\ref{gaugeanomaly}) to Fermi arcs is under current study which will be reported elsewhere. In principle, the action (\ref{weyllplus}) can be generalized to the curved space in order to search for the gravitational effects.  

We also generalized our monopole construction to any even $d+1$ dimensions (\ref{cdk}). We identified the gauge field of the Dirac monopole as a $2n-1$-rank antisymmetric tensor gauge field (\ref{agfd}). The existence of these monopoles are essential for the semiclassical chiral anomaly and the semiclassical CME. It is interesting although not surprising that the chirality-induced monopole is responsible for the anomalies even at the semiclassical level.


\subsection*{Acknowledgments}
I am grateful to \"{O}mer F. Dayi, Peter  Horv\'athy, Maxim Chernodub and Mikhail Zubkov for their comments on this work. I am also indebted to our High Energy Nuclear Physics Group at IMP for fruitful discussions. 
This work was supported by the Major State Basic Research Development Program in China (No. 2015CB856903) and the Chinese Academy of Sciences President’s International Fellowship Initiative (No. 2017PM0045).

\vspace{1cm}
\noindent
{\bf\Large{Appendix}}

\appendix

\renewcommand{\theequation}{\thesection.\arabic{equation}}
\setcounter{equation}{0}


\section{$5+1$ Dimensional Weyl Hamiltonian and Monopole }

The $5+1$ dimensional Weyl Hamiltonian ${\cal{H}}_\ssW^5$ is a $4\times 4$ matrix expressed in terms of the $5$-dimensional momentum vector $\bm{p}$ and the $3+1$ dimensional Weyl Hamiltonian ${\cal{H}}_\ssW^3$ (\ref{hw3}) as 
$$
\label{hw5}
{\cal{H}}_\ssW^5=\bm{\varSigma}\cdot\bm{p}
=\begin{pmatrix} {\cal{H}}_\ssW^3 & i(p_4+ip_5)\\
-i(p_4-ip_5) & -{\cal{H}}_\ssW^3
\end{pmatrix}.
$$   
$\varSigma_i$ are the extensions of the Pauli spin matrices:
$$
\varSigma_{a}=
\begin{pmatrix}
\sigma_{a} & 0\\
0 & -\sigma_{a}
\end{pmatrix};\ {\scriptstyle a=1,2,3},\quad
\varSigma_4=
\begin{pmatrix}
0 & i\\
-i & 0
\end{pmatrix}, \quad
\varSigma_5=
\begin{pmatrix}
0 & -1\\
-1 & 0
\end{pmatrix}.
$$

In $5+1$ dimensions (\ref{cdk}) becomes
\be
\label{dwn5}
{\cal{C}}_5= -\frac{i}{8\pi^3}\int{d^5p\ dw\ \epsilon^{ijklm}\Tr[(G^2\partial_iG^{-1})(G\partial_jG^{-1})(G\partial_kG^{-1})(G\partial_lG^{-1})(G\partial_mG^{-1})]}, 
\ee
where $i,j...=1,5$. Performing the $w$ integral and using  (\ref{dgfg}), (\ref{deriv}) we find
$$
{\cal{C}}_5=\frac{1}{8\pi^2}\int{d^5p\ \epsilon^{ijklm}\Tr[\partial_iP^+\partial_jP^+\partial_kP^+\partial_lP^+\partial_mP^+]}.
$$    
This can be written as the total derivative,
$$
{\cal{C}}_5=\frac{1}{8\pi^2}\int{d^5p\ \bm{\nabla}\cdot\bm{K}_5}, \quad K_5^i=\epsilon^{ijklm}\Tr[P^+\partial_jP^+\partial_kP^+\partial_lP^+\partial_mP^+],
$$
and using  the definition (\ref{ph}) we obtain,
\begin{eqnarray*}
K^i_5=\frac{1}{(2|\bm{p}|)^5}\epsilon^{ijklm}\Tr[{\cal{H}}^5_\ssW(\partial_j{\cal{H}}^5_\ssW)(\partial_k{\cal{H}}^5_\ssW)(\partial_l{\cal{H}}^5_\ssW)(\partial_m{\cal{H}}^5_\ssW)]
=6\frac{p^i}{2|\bm{p}|^5}.
\end{eqnarray*}
As in the $3+1$ dimensional case, we observe that (\ref{dwn5}) yields the Dirac monopole $\bm{b}_5=\frac{\bm{p}}{2|\bm{p}|^5}, \bm{\nabla}\cdot\bm{b}_5=\frac{4\pi^2}{3}\delta^5(|\bm{p}|)$. Therefore, we conclude that  the value of (\ref{dwn5}) is equal to the unit charge of this monopole:
\be
\label{mf5}
{\cal{C}}_5=\frac{3}{4\pi^2}\int d^5p\ \bm{\nabla}\cdot\bm{b}_5=1.
\ee

We can accomplish the relation between the Berry gauge field (\ref{dbgu}) and (\ref{dwn5}) by substituting (\ref{dpog}) into $K_5^i$:
\begin{eqnarray*}
K_5^i=\epsilon^{ijklm}\Tr &[&{\cal{I}}^+\partial_jU\partial_kU^{\dagger}{\cal{I}}^+\partial_lU\partial_mU^{\dagger}{\cal{I}}^+ \\ 
&+&2{\cal{I}}^+U\partial_jU^{\dagger}{\cal{I}}^+U\partial_kU^{\dagger}{\cal{I}}^+\partial_lU\partial_mU^{\dagger}{\cal{I}}^+ \\
&+&{\cal{I}}^+U\partial_jU^{\dagger}{\cal{I}}^+U\partial_kU^{\dagger}{\cal{I}}^+U\partial_lU^{\dagger}{\cal{I}}^+U\partial_mU^{\dagger}{\cal{I}}^+]\\
&=&-\frac{1}{4}\epsilon^{ijklm}\Tr[{\cal{G}}_{jk}{\cal{G}}_{lm}].
\end{eqnarray*}
Therefore, the monopole charge (\ref{dwn5}) is related to the Berry curvature:
\be
\label{wnc5}
{\cal{C}}_5=-\frac{1}{32\pi^2}\int{d^5p\ \partial_i\epsilon^{ijklm}\Tr[{\cal{G}}_{jk}{\cal{G}}_{lm}]}. 
\ee

On the other hand, by letting the domain of the integral (\ref{wnc5}) to be $B^5$, whose boundary is $S^4$, ${\cal{C}}_5$ can be written as
$$
{\cal{C}}_5=-\frac{1}{32\pi^2}\int_{\sigma_5}{d^5p\ \epsilon^{ijklm}\partial_i(\Tr[{\cal{G}}_{jk}{\cal{G}}_{lm}]})=-\frac{1}{32\pi^2}\int_{S^4}d^4p\ \epsilon^{\rm{j}\rm{k}\rm{l}\rm{m}}\Tr[{\cal{G}}_{\rm{j}\rm{k}}{\cal{G}}_{\rm{l}\rm{m}}].
$$
$\rm{j},\rm{k}...$ represent the angular coordinates of $S^4$. We conclude that the monopole charge, thus the chirality, is minus the second Chern number which is the integral of the second Chern character over $S^4$. 

Expressing the 2-form Berry field strength as ${\cal{G}}=\frac{1}{2}{\cal{G}}_{ij}dp^i\w dp^j$ and recalling (\ref{mf5}) we define the Abelian 3-form antisymmetric gauge field ${\cal{B}}_5$,  
\begin{eqnarray*}
\Tr[{\cal{G}}{\cal{G}}]=d{\cal{B}}_5.
\end{eqnarray*}
${\cal{B}}_5$ can be written explicitly as 
$$
\label{agf5}
{\cal{B}}_5=\Tr[{\cal{A}}d{\cal{A}}-\frac{2i}{3}{\cal{A}}^3], \quad {\cal{B}}^{ijk}_5=\Tr[{\cal{A}}^i\partial^j{\cal{A}}^k-\frac{2i}{3}{\cal{A}}^i{\cal{A}}^j{\cal{A}}^k],
$$
where ${\cal{A}}$ is the Berry potential (\ref{dbgu}).  This is in agreement with the general result (\ref{agfd}).


\section{Calculation of Anomaly Equations}

We will explicitly calculate anomaly equations (\ref{ncon1}) for our $S^1_{eff}$. Remaining equations (\ref{ncon2}) follow directly by changing gauge fields $A_\mu$ and $B_\mu$ and the corresponding momenta with each other. We refer to \cite{Srednicki, be} where similar calculations were carried. 

In momentum space, first equation in (\ref{ncon1}), i.e., $\partial_\mu <j^\mu_B(x)>$ becomes,
\be
\label{S1B}
A_\mu(k_1) A_\nu (k_2) [-k_1-k_2]_\rho \Pi^{\mu\nu\rho} (k_1, k_2),
\ee
where 
\be
\Pi^{\mu\nu\rho}=\int \frac{d^4p}{(2\pi)^4}\frac{\Tr\Big[\gamma^\mu \slashed{p}\gamma^\nu (\slashed{p}+\slashed{k_2})\gamma^\rho(\slashed{p}-\slashed{k_1})\gamma^5\Big]}{p^2(p+k_2)^2(p-k_1)^2},
\ee
up to constant factors.
Then, (\ref{S1B}) turns out to be,
\begin{eqnarray}
&&-A_\mu(k_1) A_\nu (k_2)\int \frac{d^4p}{(2\pi)^4}\frac{\Tr\Big[\gamma^\mu \slashed{p}\gamma^\nu (\slashed{p}+\slashed{k_2})(\slashed{k_1}+\slashed{k_2})(\slashed{p}-\slashed{k_1})\gamma^5\Big]}{p^2(p+k_2)^2(p-k_1)^2},\\
&=&-A_\mu(k_1) A_\nu (k_2)\int \frac{d^4p}{(2\pi)^4} \left\{\frac{\Tr\Big[\gamma^\mu \slashed{p}\gamma^\nu(\slashed{p}-\slashed{k_1})\gamma^5\Big]}{p^2(p-k_1)^2}-\frac{\Tr\Big[\gamma^\mu \slashed{p}\gamma^\nu (\slashed{p}+\slashed{k_2})\gamma^5\Big]}{p^2(p+k_2)^2}\right\},\quad \quad
\end{eqnarray}
where we have used $k_1+k_2=(p+k_2)+(k_1-p)$. Utilizing trace property,
\be
\Tr[\gamma^\mu \gamma^\nu \gamma^\rho\gamma^\lambda\gamma^5]=-4i\epsilon^{\mu\nu\rho\lambda},
\ee
we find
\begin{eqnarray}
&&4i \epsilon^{\mu\alpha\nu\beta}A_\mu(k_1) A_\nu (k_2)\int \frac{d^4p}{(2\pi)^4} \left\{\frac{p_\alpha (p-k_1)_\beta}{p^2(p-k_1)^2}-\frac{p_\alpha(p+k_2)_\beta}{p^2(p+k_2)^2} \right\},\\
&=&-4i \epsilon^{\mu\alpha\nu\beta}A_\mu(k_1) A_\nu (k_2)\int \frac{d^4p}{(2\pi)^4}\left\{\frac{p_\alpha {k_1}_\beta}{p^2(p-k_1)^2}+\frac{p_\alpha{k_2}_\beta}{p^2(p+k_2)^2} \right\}.
\label{B1}
\end{eqnarray}
Observe that the first term in (\ref{B1}) depends only on $p$ and $k_1$, thus any Lorentz invariant regularization of its integral necesssarily yields a term proportional to ${k_1}_\alpha {k_1}_\beta$. Similarly, second term depends only on $p$ and $k_2$, yielding ${k_2}_\alpha {k_2}_\beta$. Since antisymmetric $4-$d Levi-Civita symbol $\epsilon^{\mu\alpha\nu\beta}$ is contracted with them, the results automatically vanish. Therefore we conclude that,
\be
\label{bft1}
\partial_\mu \langle j^\mu_B(x)\rangle=0.
\ee  

Now, we consider $\partial_\mu <j^\mu_A(x)>$ for $S^1_{eff}$.  As there are 2 external legs,  $A_\mu(k_1)$ and $A_\mu(k_2)$, we will consider them separately and then sum their results. For $A_\mu(k_1)$, we need to calculate,
\begin{eqnarray}
&&A_\nu (k_2) B_\rho(-k_1-k_2) {k_1}_\mu \Pi^{\mu\nu\rho}(k_1, k_2),\\
&=&A_\nu (k_2) B_\rho(-k_1-k_2) \int \frac{d^4p}{(2\pi)^4}\frac{\Tr\Big[\slashed{k_1} \slashed{p}\gamma^\nu (\slashed{p}+\slashed{k_2})\gamma^\rho(\slashed{p}-\slashed{k_1})\gamma^5\Big]}{p^2(p+k_2)^2(p-k_1)^2},\\
&=&- A_\nu (k_2) B_\rho(-k_1-k_2) \int \frac{d^4p}{(2\pi)^4}\frac{\Tr\Big[(\slashed{p}-\slashed{k_1})\slashed{k_1} \slashed{p}\gamma^\nu (\slashed{p}+\slashed{k_2})\gamma^\rho\gamma^5\Big]}{p^2(p+k_2)^2(p-k_1)^2},
\end{eqnarray}
where we have used the cyclic property of trace operation and anticommutation relation $\{\gamma^5, \gamma^\mu\}=0$. With $k_1= p+(k_1-p)$, we find
\begin{eqnarray}
&&A_\nu (k_2) B_\rho(-k_1-k_2)\int \frac{d^4p}{(2\pi)^4}\left\{\frac{\Tr[\slashed{p}\gamma^\nu(\slashed{p}+\slashed{k_2})\gamma^\rho\gamma^5]}{p^2(p+k_2)^2} -\frac{ \Tr[(\slashed{p}-\slashed{k_1})\gamma^\nu(\slashed{p}+\slashed{k_2})\gamma^\rho\gamma^5]}{(p-k_1)^2(p+k_2)^2}\right\},\quad \quad\\
&=&4i \epsilon^{\alpha\nu\beta\rho}A_\nu (k_2) B_\rho(-k_1-k_2)\int \frac{d^4p}{(2\pi)^4}\left\{ \frac{(p-k_1)_\alpha (p+k_2)_\beta}{(p+k_2)^2(p-k_1)^2}-\frac{p_\alpha(p+k_2)_\beta}{p^2(p+k_2)^2}\right\}.
\label{a1simplify}
\end{eqnarray}
Now, second part of (\ref{a1simplify}) vanishes since any gauge invariant regularization yields $\epsilon^{\alpha\nu\beta\rho}{k_2}_\alpha {k_2}_\beta=0$.  First term,
\begin{eqnarray}
&&4i \epsilon^{\alpha\nu\beta\rho}A_\nu (k_2) B_\rho(-k_1-k_2)\int \frac{d^4p}{(2\pi)^4} \frac{(p-k_1)_\alpha (p+k_2)_\beta}{(p+k_2)^2(p-k_1)^2},\\
&=&-4i \epsilon^{\alpha\nu\beta\rho}A_\nu (k_2) B_\rho(-k_1-k_2)[k_1+k_2]_\alpha \int \frac{d^4p}{(2\pi)^4} \frac{(p+k_2)_\beta}{(p+k_2)^2(p-k_1)^2},\quad
\end{eqnarray}
seems to be vanishing if we could shift the integral variable, e.g., $p\to p-k_2$. However, such a naive shift is problematic since our integral is linearly divergent. We define and expand it as
\begin{subequations}
\label{surface}
\begin{align}
&\int \frac{d^4p}{(2\pi)^4} \frac{(p+k_2)_\beta}{(p+k_2)^2(p-k_1)^2}\equiv \int \frac{d^4p}{(2\pi)^4} f(p+k_2)_\beta,\\
&\int \frac{d^4p}{(2\pi)^4} f(p+k_2)_\beta=\int \frac{d^4p}{(2\pi)^4} \left\{f(p)_\beta+k_2^\delta \frac{\partial f(p)_\beta}{\partial p^\delta}+... \right\},\\
&\int \frac{d^4p}{(2\pi)^4} f(p)_\beta+ i k_2^\delta\lim_{p\to\infty}\int \frac{dS_\delta}{(2\pi)^4} f(p)_\beta
\label{infexpansion}
\end{align}
\end{subequations}
where $dS_\delta= p^2 p_\delta d\Omega$ is the surface area element and $d\Omega$ is the differential solid angle. In the second line, we have ignored higher order derivatives  since they fall off rapidly at infinity, while in the third line we performed a Wick rotation, transforming the integral to the Euclidean space. Surface integral can be calculated as
\begin{eqnarray}
\label{surfacer}
 i k_2^\delta\lim_{p\to\infty}\int \frac{dS_\delta}{(2\pi)^4}\frac{p_\beta}{p^2(p-k_1-k_2)^2}=ik_2^\delta\frac{2\pi^2}{(2\pi)^4}\frac{\eta_{\delta\beta}}{4}=\frac{i{k_2}_\beta}{32\pi^2}.
\end{eqnarray}
Using the expansion (\ref{infexpansion}) and (\ref{surfacer}), we obtain two integrals. One of them vanishes as $\epsilon^{\alpha\nu\beta\rho}(k_1+k_2)_\alpha (k_1+k_2)_\beta=0$. It is the surface integral that causes the anomaly and we find the non-conservation for $A_\mu(k_1)$ in momentum space as
\be
\label{nca1}
A_\nu (k_2) B_\rho(-k_1-k_2) {k_1}_\mu \Pi^{\mu\nu\rho}(k_1, k_2)=
\frac{1}{8\pi^2}\epsilon^{\alpha\nu\beta\rho} A_\nu (k_2) B_\rho(-k_1-k_2){k_1}_\alpha {k_2}_\beta.
\ee

Similarly, we can obtain the nonconservation for the $A_\nu(k_2)$ leg as,
\begin{eqnarray}
&&A_\mu (k_1) B_\rho(-k_1-k_2) {k_2}_\nu \Pi^{\mu\nu\rho}(k_1, k_2),\\
&=&A_\mu (k_1) B_\rho(-k_1-k_2) \int \frac{d^4p}{(2\pi)^4}\frac{\Tr\Big[\gamma^\mu \slashed{p}\slashed{k_2} (\slashed{p}+\slashed{k_2})\gamma^\rho(\slashed{p}-\slashed{k_1})\gamma^5\Big]}{p^2(p+k_2)^2(p-k_1)^2}.
\end{eqnarray}
 With $k_2= p+(k_2-p)$, we find
\begin{eqnarray}
&&-A_\mu (k_1) B_\rho(-k_1-k_2)\int \frac{d^4p}{(2\pi)^4}\left\{\frac{\Tr[\gamma^\mu(\slashed{p}+\slashed{k_2})\gamma^\rho(\slashed{p}-\slashed{k_1})\gamma^5]}{(p-k_1)^2(p+k_2)^2} -\frac{ \Tr[\gamma^\mu\slashed{p}\gamma^\rho(\slashed{p}-\slashed{k_1})\gamma^5]}{(p-k_1)^2p^2}\right\},\quad \quad\\
&=&4i \epsilon^{\mu\alpha\rho\beta}A_\mu (k_1) B_\rho(-k_1-k_2)\int \frac{d^4p}{(2\pi)^4}\left\{\frac{(p+k_2)_\alpha (p-k_1)_\beta}{(p+k_2)^2(p-k_1)^2}-\frac{p_\alpha(p-k_1)_\beta}{p^2(p-k_1)^2}\right\}.
\label{a2simplify}
\end{eqnarray}
Second part of (\ref{a2simplify}) vanishes since any gauge invariant regularization must yield $\epsilon^{\mu\alpha\rho\beta}{k_1}_\alpha {k_1}_\beta=0$.  First term becomes
\begin{eqnarray}
&&4i \epsilon^{\mu\alpha\rho\beta}A_\mu (k_1) B_\rho(-k_1-k_2)\int \frac{d^4p}{(2\pi)^4} \frac{(p+k_2)_\alpha (p-k_1)_\beta}{(p+k_2)^2(p-k_1)^2},\\
&=&-4i \epsilon^{\mu\alpha\rho\beta}A_\mu (k_1) B_\rho(-k_1-k_2)[k_1+k_2]_\beta\int \frac{d^4p}{(2\pi)^4}\frac{(p+k_2)_\alpha}{(p+k_2)^2(p-k_1)^2}.\quad
\end{eqnarray}
This seems to be vanishing with the naive shift $p\to p-k_2$, however we have to deal with the boundary terms again. Repeating the steps carried in (\ref{surface}) and (\ref{surfacer}), we find the non-conservation for $A_\nu(k_2)$ in momentum space as
\be
\label{nca2}
A_\mu (k_1) B_\rho(-k_1-k_2) {k_2}_\nu \Pi^{\mu\nu\rho}(k_1, k_2)=
\frac{1}{8\pi^2}\epsilon^{\mu\alpha\rho\beta} A_\mu (k_1) B_\rho(-k_1-k_2){k_1}_\beta {k_2}_\alpha.
\ee
A Fourier transformation of the sum (\ref{nca1})+(\ref{nca2}) results in the non-conservation of $A-$current as
\be
\label{aft1}
\partial_\mu \langle j^\mu_A(x)\rangle= \frac{e^2}{24\pi^2}\epsilon^{\mu\nu\rho\lambda}\partial_\mu B_\nu \partial_\rho A_\lambda.
\ee

(\ref{bft1}) and (\ref{aft1}) provides our (\ref{ncon1}). Following the same lines one can calculate the remaining nonconservation equations in (\ref{ncon2}).

\newpage

\newcommand{\PRL}{Phys. Rev. Lett. }
\newcommand{\PRB}{Phys. Rev. B }
\newcommand{\PRD}{Phys. Rev. D }

\end{document}